\documentclass[dvipsnames]{article}

\usepackage[final]{openbmb_2025}

\usepackage{amsmath}
\usepackage{amssymb}
\usepackage{mathtools}
\usepackage{amsthm}
\usepackage{multirow}
\usepackage{makecell}
\usepackage{subcaption}
\usepackage{wrapfig}
\usepackage{graphicx}

\usepackage{pifont}
\usepackage{fvextra}
\usepackage{xspace}

\usepackage[utf8]{inputenc} % allow utf-8 input
\usepackage[T1]{fontenc}    % use 8-bit T1 fonts
\usepackage{url}            % simple URL typesetting
\usepackage{booktabs}       % professional-quality tables
\usepackage{amsfonts}       % blackboard math symbols
\usepackage{nicefrac}       % compact symbols for 1/2, etc.
\usepackage{microtype}      % microtypography

\usepackage{color, soul}
\usepackage[most,breakable]{tcolorbox}
\usepackage{textgreek}

\usepackage[labelfont=bf]{caption}
\usepackage{enumitem}
\usepackage{sectsty}
\usepackage{pgfplots}
\usepackage{fancyhdr}
\usepackage{xcolor} % 引入xcolor宏包
\usepackage{bbding}
\renewcommand{\headrulewidth}{1pt}
\makeatletter
\def\headrule{{\if@fancyplain\let\headrulewidth\plainheadrulewidth\fi
\hrule\@height\headrulewidth\@width\textwidth \vskip-\headrulewidth}}
\makeatother

\definecolor{BMBDarkBlue}{HTML}{315EFE}
\definecolor{BMBLightBlue}{HTML}{00D3ED}
\usepackage[colorlinks, linkcolor=BMBDarkBlue, anchorcolor=BMBDarkBlue, citecolor=BMBDarkBlue, urlcolor=BMBDarkBlue]{hyperref}
\usepackage{cleveref}
\usepackage{threeparttable}

\sectionfont{\color{BMBDarkBlue}\fontfamily{zi4}\selectfont}
\subsectionfont{\color{BMBDarkBlue}\fontfamily{zi4}\selectfont}
\subsubsectionfont{\color{BMBDarkBlue}\fontfamily{zi4}\selectfont}

\newtcolorbox{mytheorem}{
  colback=gray!5,       % BackGround Color
  colframe=gray!80,     % Frame Color
  boxrule=0.5pt,        % Frame Width
  arc=4pt,              % Fillet Radius
  left=4pt,             % Left Margin
  right=4pt,            % Right Margin
  top=4pt,              % Top Margin
  bottom=4pt,           % Bottom Margin
}

\newcommand{\fancyheadname}{\textit{\textbf{\modelname{}}}}
\title{\modelname{}: Tokenizer-Free TTS for Context-Aware Speech Generation and True-to-Life Voice Cloning}

\author{%
\\
\textbf{\large{VoxCPM Team}}
\vspace{0em}
}

\def\huggingface{\raisebox{-1.5pt}{\includegraphics[height=1.05em]{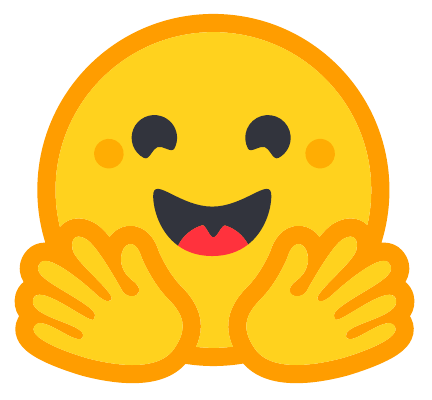}}}
\def\github{\raisebox{-1.5pt}{\includegraphics[height=1.05em]{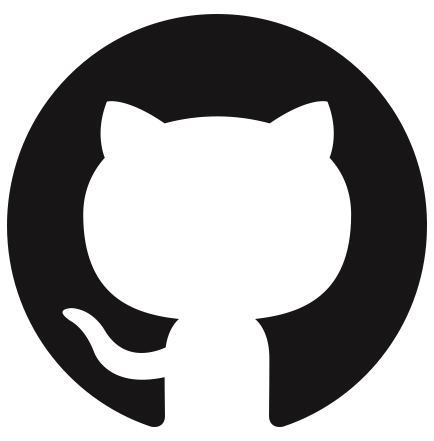}}}
\newcommand{\modelname}[0]{VoxCPM}

\usepackage{geometry}
\usepackage{algorithm}
\usepackage{algpseudocode}
\usepackage{setspace}

\makeatletter
\renewcommand{\ALG@beginalgorithmic}{\small}
\makeatother



\begin{document}
\maketitle
\thispagestyle{fancy}
\begin{abstract}
Generative models for speech synthesis face a fundamental trade-off: discrete tokens ensure stability but sacrifice expressivity, while continuous signals retain acoustic richness but suffer from error accumulation due to task entanglement. 
This challenge has driven the field towards multi-stage pipelines that rely on pre-trained speech tokenizers, but these create a semantic-acoustic divide, limiting holistic and expressive speech generation.  
We resolve these dilemma through hierarchical semantic-acoustic modeling with semi-discrete residual representations and present a novel tokenizer-free TTS model--\modelname{}.
Our framework introduces a differentiable quantization bottleneck that induces natural specialization: 
a Text-Semantic Language Model (TSLM) generates semantic-prosodic plans, while a Residual Acoustic Model (RALM) recovers fine-grained acoustic details.
This hierarchical semantic-acoustic representation guides a local diffusion-based decoder to generate high-fidelity speech latents. 
Critically, the entire architecture is trained end-to-end under a simple diffusion objective, eliminating dependency on external speech tokenizers. 
Trained on a massive 1.8 million hours of bilingual corpus, our \modelname{}-0.5B model achieves state-of-the-art zero-shot TTS performance among open-source systems, demonstrating that our approach delivers expressive and stable synthesis. 
Besides, VoxCPM shows the capability to comprehend text to infer and generate appropriate prosody and style, delivering speech with context-aware expressiveness and natural flow. 
To facilitate community-driven research and development, \modelname{} is publicly accessible under Apache 2.0.
\end{abstract}

\newpage
{
  \hypersetup{linkcolor=RoyalBlue, linktoc=page}
  \tableofcontents
}

\newpage

\section{Introduction}
The pursuit of modern text-to-speech (TTS) systems has evolved beyond intelligibility toward the synthesis of genuinely human-like audio, capable of conveying subtle emotions, speaker identity, and contextual nuances \citep{shen2018natural, ping2017deep, renfastspeech, li2019neural}. 
This leap is critical for applications like empathetic virtual assistants and immersive digital avatars, and hinges on a core technical challenge: simultaneously capturing the fine-grained acoustic details that define vocal richness and the long-range semantic structures governing intelligibility and natural prosody.

Inspired by the success of large language models (LLMs), a dominant paradigm frames TTS as a sequence modeling task over discrete tokens from pre-trained neural audio codecs (e.g., EnCodec \citep{defossez2022high}).
Autoregressively or Non-autoregressively predicting these tokens from text or phonemes \citep{borsos2023audiolm, kharitonov2023speak, chen2025neural, wangmaskgct, peng2024voicecraft} offers excellent scalability and in-context learning capabilities. 
However, this approach faces a fundamental "quantization ceiling", as the compression process irreversibly discards subtle acoustic details.
To mitigate this quality loss, state-of-the-art TTS systems \citep{du2024cosyvoice1, du2024cosyvoice2, du2025cosyvoice3, zhou2025indextts2, casanova2024xtts} adopt multi-stage hybrid pipelines.
Here, an LLM generates discrete tokens which condition a separate diffusion-based decoder.
While improving fidelity, this solution creates a stark semantic-acoustic divide: the LLM operates in an abstract, discrete space unaware of acoustic reality, while the diffusion model performs local refinement without high-level context. 
This fragmentation prevents end-to-end optimization and limits holistic, expressive and context-aware speech synthesis.

Alternatively, other approaches directly model continuous speech representations to avoid quantization loss. 
Early systems like Tacotron 2 \citep{shen2018natural} and more recent models such as MELLE \citep{meng2024autoregressive} generate mel-spectrograms autoregressively. 
However, predicting continuous targets under standard regression losses often yields over-smoothed and low-diversity outputs. 
To address this, recent innovations have explored replacing the regression objective with a denoising process to model the distribution of the next continuous representations, spanning both non-autoregressive paradigms \citep{shen2023naturalspeech, le2023voicebox,chen2024f5} and autoregressive methods\citep{li2024autoregressive, jia2025ditar, peng2025vibevoice}. 
Among these, autoregressive approaches have often demonstrated superior performance in capturing natural prosody and expressive variation.
This innovation successfully enhances the detail and diversity of generated continuous representations. 
However, a more fundamental issue persists: in a fully continuous autoregressive model, the tasks of high-level semantic-prosodic planning and fine-grained acoustic rendering are conflated within a single learning objective. 
The model is forced to simultaneously solve two disparate tasks—requiring different inductive biases—in a continuous output space. 
This entanglement presents a significant challenge to the modeling capacity of a single LLM, as it must learn to be both a global planner and a local renderer without an inherent architectural bias to separate these functions.
We argue that this conflation is a root cause of instability. 
The model's focus is inevitably pulled towards fitting low-level acoustic textures, which compromises its ability to maintain high-level semantic coherence, leading to the well-known problem of error accumulation over long sequences \citep{pasini2024continuous}.

In this work, we introduce a tokenizer-free, end-to-end framework that resolves this trade-off through hierarchical semantic-acoustic modeling with semi-discrete residual representations and present a novel TTS model--\modelname{}.
Our key insight is that holistic and expressive speech synthesis requires explicit architectural separation between semantic-prosodic planning and acoustic rendering, yet should remain within a cohesive, end-to-end trainable system.
The core innovation is a differentiable Finite Scalar Quantization (FSQ) \citep{mentzerfinite} bottleneck that induces natural specialization:
(1) a Text-Semantic Language Model (TSLM) generates semantic-prosodic plans stabilized through quantization, focusing on linguistically meaningful patterns;
and (2) a Residual Acoustic Language Model (RALM) recovers fine-grained details lost during quantization, specializing in acoustic refinement.
This hierarchical design enables each component to excel at its respective role while maintaining differentiability, and both of them will be used to guide a local diffusion decoder to generate high-fidelity speech latents.
Critically, the entire hierarchical model is trained end-to-end under a simple diffusion objective, seamlessly integrating planning and rendering without pre-trained tokenizers.
Trained on a massive 1.8 million hours of bilingual corpus, our \modelname{}-0.5B model achieves state-of-the-art zero-shot TTS performance among open-source systems, demonstrating that our approach delivers expressive and stable synthesis. 
Our main contributions are as follows:
\vspace{-2mm}
\begin{itemize}[itemsep=0.2em, leftmargin=2em]
    \item We propose an end-to-end hierarchical architecture that introduces an internal semi-discrete bottleneck to resolve the expressivity-stability trade-off. This mechanism implicitly addresses task entanglement in continuous models by inducing a beneficial separation between semantic-prosodic planning and fine-grained acoustic modeling within a single, unified framework.
    \item We introduce a residual learning strategy that, in conjunction with the bottleneck, enables a holistic yet specialized modeling process. Unlike fragmented multi-stage pipelines, our approach achieves functional separation without architectural fragmentation, simplifying the training pipeline and eliminating dependency on external speech tokenizers.
    \item We demonstrate the efficacy of our approach through large-scale training on a massive 1.8 million hours of bilingual speech. The resulting model, VoxCPM-0.5B, achieves state-of-the-art zero-shot TTS performance among open-source systems with a Real-Time Factor (RTF) as low as 0.17 on a consumer-grade NVIDIA RTX 4090 GPU, validating its practical strength.
    \item We provide extensive ablation studies that conclusively validate the semi-discrete residual representations as the crucial component for robust, expressive, and l context-aware synthesis. Besides, we release the codes and models publicly to support community development and future research.
\end{itemize}

\section{Related Work}
\subsection{Discrete Token-Based TTS} 
The discrete token paradigm has emerged as a dominant approach in modern TTS, leveraging the success of large language models. This method converts speech into discrete representations using neural audio codecs such as EnCodec \citep{defossez2022high} and DAC \citep{kumar2023high} through residual vector quantization (RVQ).
AudioLM \citep{borsos2023audiolm} and VALL-E \citep{chen2025neural} pioneered this direction by framing audio generation and TTS as an autoregressive sequence prediction task over discrete acoustic tokens. Subsequent developments include SoundStorm \citep{borsos2023soundstorm}, which introduced non-autoregressive generation for improved efficiency, and Spear-TTS \citep{kharitonov2023speak}, which focused on multilingual capabilities with minimum supervision. 
Besides, VoiceCraft \citep{peng2024voicecraft} and XTTS \citep{casanova2024xtts} further advanced zero-shot TTS with in-context learning.

Recent advancements have focused on enhancing the scalability, controllability and zero-shot adaptation. 
CosyVoice \citep{du2024cosyvoice1} proposed supervised semantic tokens for improved zero-shot performance, while its successors, 
%CosyVoice 2 \citep{du2024cosyvoice2} and CosyVoice 3 \citep{du2025cosyvoice3}, 
CosyVoice 2 and 3 \citep{du2024cosyvoice2, du2025cosyvoice3}
incorporated text-based LLM initialization, streaming synthesis, and large-scale training data for human-parity quality, low latency and in-the-wild scenarios.
IndexTTS \citep{deng2025indextts} and IndexTTS2 \citep{zhou2025indextts2} introduced precise duration and emotion control in autoregressive token generation, enabling applications with strict timing and expressivity requirements. SparkTTS \citep{wang2025spark} utilized single-stream decoupled speech tokens for modeling efficiency, and FireRedTTS \citep{guo2024fireredtts} along with its update FireRedTTS-2 \citep{xie2025fireredtts} established frameworks for industry-level generative speech, including long-form multi-speaker dialogue.   
Openaudio-s1 \citep{openaudios1} used dual AR architecture and online Reinforcement Learning from Human Feedback (RLHF) to improve expressiveness and instruction-following capabilities.
Higgs Audio v2 \citep{higgsaudio} proposed a unified audio tokenizer captures
both semantic and acoustic features, and  pretrained on over 10 million hours of audio data, providing a powerful foundation model. 
Despite these progresses, discrete approaches suffer from inherent quantization artifacts, limiting acoustic fidelity and prompting hybrid solutions.

\vspace{-0.2cm}
\subsection{Continuous Representation TTS}  
To circumvent quantization losses in discrete models, continuous representation approaches directly model speech features such as mel-spectrograms or audio latents.
Early systems like Tacotron 2 \citep{shen2018natural} established the encoder-decoder framework for text-to-mel mapping, while FastSpeech \citep{renfastspeech} introduced explicit duration modeling for alignment stability.
Inspired from VALL-E,  MELLE \citep{meng2024autoregressive} autoregressively generated continuous mel-spectrogram frames directly from text condition, and incorporated variational inference to facilitate sampling mechanisms.
Recent developments have integrated diffusion processes to enhance detail and diversity. Non-autoregressive models like NaturalSpeech 2 \citep{shen2023naturalspeech} and VoiceBox \citep{le2023voicebox} apply diffusion directly on continuous representations.
F5-TTS \citep{chen2024f5} advanced flow-matching for efficient synthesis. 
Autoregressive paradigms, often superior in prosody and variation, additionally possess the capability for streaming synthesis.
Innovations like ARDiT \citep{li2024autoregressive} use an autogressive diffusion transformer for TTS, unifying semantic coherence and acoustic naturalness via parameter sharing. 
DiTAR \citep{jia2025ditar} extended this with a patch-based design: a causal LM for inter-patch stability and a bidirectional local diffusion transformer for intra-patch refinement. 
VibeVoice \citep{peng2025vibevoice} employed next-token diffusion for long-form multi-speaker synthesis. 
Besides, recent models such as CLEAR \citep{wu2025clear} and FELLE \citep{wang2025felle} focus on latent autoregressive modeling with token-wise coarse-to-fine hierarchies, while MELA-TTS \citep{an2025mela} and KALL-E \citep{zhu2024autoregressive} combine joint transformer-diffusion with next-distribution prediction for improved efficiency and quality. Despite these advances, continuous models often entangle high-level semantic planning with low-level acoustic rendering, leading to instability in long sequences without explicit separation.

\vspace{-0.2cm}
\subsection{Hierarchical and Residual Modeling in TTS}
Hierarchical and residual approaches decompose TTS into layered tasks to balance stability and expressivity.
HierSpeech++ \citep{lee2025hierspeech++} employed variational inference for semantic-acoustic mapping. 
HALL-E \citep{nishimurahall} uses hierarchical neural codecs with LLMs for minute-long synthesis. MARS6 \citep{baas2025mars6} builds robust encoder-decoder transformers with hierarchical tokens. DiffStyleTTS \citep{liu2024diffstyletts} applies diffusion for hierarchical prosody modeling. HAM-TTS \citep{wang2024ham} introduces hierarchical acoustic modeling with data augmentation for zero-shot TTS. 
QTTS \citep{han2025quantize} features hierarchical parallel architectures for residually quantized codes. 
In song generation, LeVo \citep{lei2025levo} likewise introduced a hierarchical framework using two decoder-only transformers for layered modeling of mixed and separated part in a song, achieving enhanced generation quality.
These methods address flaws in prior paradigms: implicit designs lack regulated bottlenecks, tokenizer-dependent models suffer discrete losses, and fragmented stages hinder end-to-end optimization. However, few fully integrate explicit residual designs with semi-discrete bottlenecks in a unified framework, as proposed in our work, to achieve implicit disentanglement without external dependencies.

\section{Methodology}
\begin{figure}[h]
\begin{center}
\includegraphics[width=1.0\textwidth]{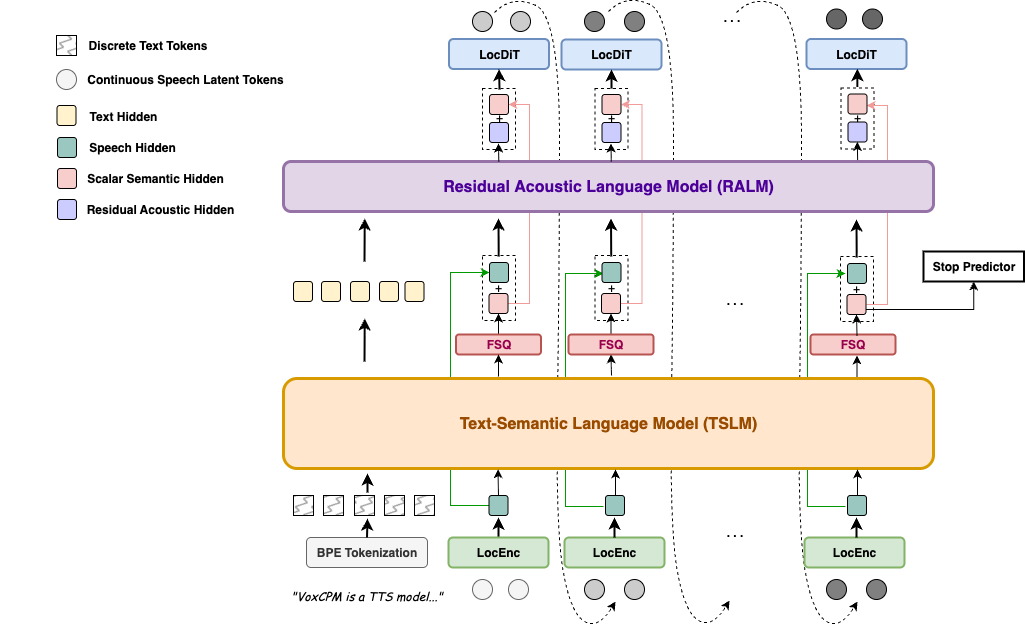}
\end{center}
\caption{Overall architecture of VoxCPM. The model hierarchically generates speech by first processing audio latents through a LocEnc, then producing a semi-discrete speech skeleton with the TSLM and FSQ, refining acoustic details with the RALM, and finally generating high-fidelity latent output with the LocDiT.}
\label{fig:architecture}
\end{figure}

\subsection{Core Design Motivation}
\label{sec:motivation}
Generative speech synthesis faces a fundamental tension between expressivity and stability.
Discrete tokenization methods (e.g., speech tokenizers with language models) ensure stable autoregressive generation but irreversibly discard fine-grained acoustic details through quantization. 
Continuous approaches preserve full fidelity but suffer from error accumulation in long sequences due to information entanglement, often leading to catastrophic failure in intelligibility.

Critically, we identify a key limitation in existing discrete tokenization approaches: methods that directly use FSQ or VQ to obtain discrete codebooks for language modeling face an inherent scalability challenge. As the dimensionality increases to capture richer acoustic information, the codebook size grows exponentially, creating an unmanageably large and sparse vocabulary that language models struggle to predict accurately.

We hypothesize that an effective solution should \textbf{structurally separate} the modeling of stable semantic-prosodic content from fine-grained acoustic details while maintaining differentiability for end-to-end training. 
Our key insight is to introduce a \textbf{differentiable quantization bottleneck} that naturally induces this separation through scalar quantization, splitting information into a discrete-like skeleton for content stability and continuous residual components for detail expressivity.

Unlike multi-stage TTS systems composed of seperate LM and diffusion that treat quantization as a means to obtain discrete prediction targets, our approach uses quantization solely as a regularization mechanism to constrain the hidden state space. 
This distinction allows us to avoid the vocabulary explosion problem while still benefiting from the stabilizing effects of discrete representations.

\subsection{Model Overview}
\label{sec:overview}

VoxCPM employs a hierarchical autoregressive architecture that generates sequences of continuous speech latents $\mathbf{Z} = \{\mathbf{z}_1, ..., \mathbf{z}_M\}$ conditioned on input text tokens $\mathbf{T} = \{t_1, ..., t_N\}$,
where each $\mathbf{z}_i \in \mathbb{R}^{P \times D}$ represents a patch of $P$ frames with $D$-dimensional VAE latent vectors. 
The generation process follows:
\begin{equation}
    p(\mathbf{Z}|\mathbf{T}) = \prod_{i=1}^M p(\mathbf{z}_i | \mathbf{T}, \mathbf{Z}_{<i})
\end{equation}

The core innovation lies in our hierarchical conditioning mechanism with residual representation learning.
It is made up of a local audio encoder (LocEnc), a text-semantic language model (TSLM), a residual acoustic language model (RALM) and a local diffusion transformer decoder (LocDiT). A stop predictor is attached to the output of the TSLM to determine the endpoint of generation. 
As shown in Figure~\ref{fig:architecture}, each patch generation involves:
\begin{equation}
    \mathbf{z}_i \sim \text{LocDiT}(\mathbf{h}_i^{\text{final}}), \quad 
    \mathbf{h}_i^{\text{final}} = \underbrace{\text{FSQ}(\text{TSLM}(\mathbf{T}, \mathbf{E}_{<i}))}_{\text{stable skeleton}} + \underbrace{\text{RALM}(\cdot)}_{\text{residual details}}
\end{equation}
where $\mathbf{E}_{<i} = \text{LocEnc}(\mathbf{Z}_{<i})$ represents historical audio context aggregated by a lightweight LocEnc that compresses VAE latent patches into compact acoustic embeddings.
The hierarchical backbone produces a conditioning signal $\mathbf{h}_i^{\text{final}}$ that encapsulates both semantic content from TSLM (with FSQ) and acoustic details from RALM. 
This signal guides the LocDiT to generate the current latent patch $z_i$ through a denoising diffusion process.
The entire model is trained end-to-end with gradients flowing through all components, including the FSQ bottleneck via straight-through estimation, ensuring coordinated optimization toward holistic speech synthesis.

\subsection{Hierarchical Semantic-Acoustic Modeling}
\label{sec:hierarchical}

Our hierarchical modeling approach is designed to implicitly separate semantic-prosodic planning from fine-grained acoustic synthesis, addressing the fundamental stability-expressivity trade-off through structured representation learning.
  
\subsubsection{Text-Semantic Language Model (TSLM)}
\label{sec:tslm}

The Text-Semantic Language Model forms the main part of our hierarchical architecture, responsible for capturing high-level linguistic structure and generating contextually appropriate speech patterns. 
Unlike conventional TTS systems that typically operate on phoneme sequences, our approach leverages a pre-trained text language model (MiniCPM-4~\citep{team2025minicpm4}) as its initial backbone, enabling richer contextual understanding and more natural prosody prediction directly from raw text.
Specifically, we employ character-level segmentation for Chinese BPE Tokenizer to mitigate the vocabulary sparsity issue in TTS tasks.
By processing both text tokens and historical audio context, the TSLM learns to generate semantic content and prosodic structure that evolve naturally throughout an utterance, reflecting the underlying linguistic meaning rather than simply mapping phonemes to acoustic features.
The TSLM produces continuous semantic-prosodic representations that encode both the content to be spoken and how it should be prosodically realized, serving as input to the subsequent quantization stage.

\subsubsection{Semi-Discrete Representation Learning via FSQ}
\label{sec:fsq}

At the core of our approach lies the Finite Scalar Quantization (FSQ) layer, which projects the continuous hidden states from the TSLM onto a structured lattice to create a semi-discrete representation. 
The FSQ operation transforms each dimension of the continuous vector through a deterministic scalar quantization:
\begin{equation}
\mathbf{h}_{i,j}^{\text{FSQ}} = \Delta \cdot \text{clip}\left( \text{round}\left( \frac{\mathbf{h}_{i,j}^{\text{TSLM}}}{\Delta} \right), -L, L \right)
\end{equation}
where $\Delta$ is the quantization step size, $L$ is the clipping range, and $\text{round}$ maps values to discrete levels. 
This transformation creates a structured discrete representation while maintaining differentiability through the straight-through estimator during backward passes.

The FSQ layer acts as a bottleneck, analogous to the first layer of Residual Vector Quantization (RVQ), which captures a coarse semantic-prosodic skeleton (e.g., content, intonation patterns). 
We term this representation ``semi-discrete" as it employs a significantly larger dimensionality than standard FSQ to ensure sufficient informational capacity.
Unlike RVQ, where the first layer is a prediction target and subsequent layers model finer details, our FSQ bottleneck serves as an intermediate, differentiable inductive bias  within the continuous data flow. 
It encourages the model to prioritize modeling stable, high-level components (the semantic-prosodic skeleton) by providing a clear learning signal for what information should be preserved through the bottleneck. 
This structured approach mitigates error accumulation by reducing the modeling burden on the TSLM, allowing it to focus on the major components of the speech.

\subsubsection{Residual Acoustic Modeling}
\label{sec:ralm}

To recover the fine-grained acoustic information attenuated by quantization, we introduce the Residual Acoustic Language Model (RALM). 
This module specializes in reconstructing those subtle vocal characteristics that conventional discrete methods sacrifice for stability. 
It processes the quantization residuals along with contextual information to recover speaker identity, spectral fine structure, and micro-prosodic variations:
\begin{equation}
    \mathbf{h}_i^{\text{residual}} = \text{RALM}( \mathbf{H_{\text{text}}^{\text{TSLM}}}, \mathbf{H}_{<i}^{\text{FSQ}} \oplus \mathbf{E}_{<i})
\end{equation}
Here, the RALM conditions its predictions on both the TSLM hidden states of the text part $\mathbf{H_{\text{text}}^{\text{TSLM}}}$, the semi-discrete representation of speech part $\mathbf{H}_{<i}^{\text{FSQ}}$, and the historical acoustic embeddings $\mathbf{E}_{<i}$.
This residual learning approach  creates a natural division of labor: the TSLM+FSQ pathway focuses on content stability and prosodic coherence, while the RALM pathway specializes in acoustic expressivity and speaker characteristics.

The final combined representation $\mathbf{h}_i^{\text{final}} = \mathbf{h}_i^{\text{FSQ}} + \mathbf{h}_i^{\text{residual}}$ thus encapsulates both semantic stability and acoustic expressivity, creating a comprehensive signal that guides the subsequent local diffusion process.

\subsubsection{Local Diffusion Transformer  Decoder}
\label{sec:locdit}

The Local Diffusion Transformer (LocDiT) serves as our high-fidelity synthesis module, generating continuous latent patches conditioned on the hierarchical representation $\mathbf{h}_i^{\text{final}}$ produced by the preceding modules.
Following DiTAR \citep{jia2025ditar}, we employ a bidirectional Transformer architecture that enables full receptive field modeling within each patch. 
To enhance generation consistency, we incorporate the previous patch $\mathbf{z}_{i-1}$ as additional conditioning context, which has been empirically validated to significantly improve output quality by framing the task as outpainting rather than independent patch generation.
Besides, we mask the LM guidance in LocDiT condition with a specific probability ratio, for enabling classifier-free guidance (CFG) during inference.

\subsection{Training Objective}
The entire model is trained end-to-end using a flow-matching objective that directly optimizes the quality of the generated speech latents. We adopt the conditional flow-matching formulation for its training stability and sampling efficiency:
\begin{equation}
\mathcal{L}_{\text{FM}} = \mathbb{E}{t, \mathbf{z}_i^0, \boldsymbol{\epsilon}} \left[ | \mathbf{v}_{\theta}(\mathbf{z}_i^t, t, \mathbf{h}_i^{\text{final}}, \mathbf{z}_{i-1}) - \frac{d}{dt}(\alpha_t \mathbf{z}_i^0 + \sigma_t \boldsymbol{\epsilon}) |^2 \right]
\end{equation}
where $\mathbf{z}_i^t = \alpha_t \mathbf{z}_i^0 + \sigma_t \boldsymbol{\epsilon}$ is the noisy latent at time $t$, with $\boldsymbol{\epsilon} \sim \mathcal{N}(0, I)$, and $\mathbf{v}_{\theta}$ is the velocity field predicted by the LocDiT.

Simultaneously, a binary classification loss is applied to train the model to predict the end of a speech sequence:
\begin{equation}
\mathcal{L}_{\text{Stop}} = \mathbb{E}_{i \sim \text{sequence}} \left[ \text{BCE}\left(s_{\theta}(\mathbf{h}_i^{\text{FSQ}}), \mathbb{1}[\text{token } i \text{ is the last}]\right) \right]
\end{equation}
where $s_{\theta}$ is a stop-logit projection layer, and BCE denotes the binary cross-entropy loss.

The gradients from this loss are backpropagated through the entire autoregressive hierarchy, including the FSQ layer (via straight-through estimation), the TSLM and the LocEnc. This end-to-end optimization 
under the combined objective $\mathcal{L} = \mathcal{L}_{\text{FM}} + \lambda \mathcal{L}_{\text{Stop}}$ 
allows each component to learn its specialized role—semantic planning, stabilization, and acoustic refinement—in a coordinated manner, guided by the unified objective of accurately modeling the continuous speech latents.

\subsection{Causal Audio VAE}
\label{sec:vae}

To enable efficient streaming synthesis, we employ a causal Variational Autoencoder that operates in a computationally efficient latent space. 
VAE is trained separately using a composite objective that combines reconstruction loss in the Mel-spectrogram domain, adversarial training with multi-period and multi-scale discriminators, and a minimal KL-divergence term to regularize the latent space.
The use of a latent space rather than raw audio waveforms significantly reduces computational requirements while preserving perceptual quality.
The causal nature of the VAE ensures that both encoding and decoding operations can be performed in a streaming fashion, making the entire system suitable for real-time applications where low latency is critical.

Specifically, the Audio VAE operates continuous speech tokens at a 25 Hz frame rate.
The VAE's architecture is similar to DAC~\citep{kumar2023high}, with both its encoder and decoder implemented using stacked Causal Convolutional Networks (Causal CNNs). 
For 16 kHz single-channel audio, the encoder achieves a 640x downsampling factor through a series of strided convolutions with a stride sequence of [2, 5, 8, 8], compressing the audio into a 25 Hz latent representation. 
The decoder then reconstructs the original waveform by upsampling from this latent representation. 
The training objectives consist of an adversarial (GAN) loss, a Mel-spectrogram loss, and a KL divergence loss, with the latter's weight set to a very small value $5e-5$.

\section{Experiments and Results}
\subsection{Experimental Setup}
\textbf{Datasets}
We conducted experiments on two primary datasets:
(1) \textbf{Large-scale Bilingual Corpus}: To explore the best performance, we collected an internal large-scale, bilingual dataset totaling on a massive 1.8 million hours, mainly comprising of Chinese and English speech. 
The raw audio was sourced from a diverse set of domains, including audiobooks, podcasts, interviews, and broadcast dramas. 
To enhance model robustness and enable advanced functionalities such as pronunciation correction, we further constructed some specialized training samples by applying data augmentation techniques, including random phoneme replacement on the transcriptions.
All audio was resampled to 16kHz mono, processed with source separation, voice activity detection (VAD), and automatic speech recognition (ASR) system to obtain text-audio alignment.
(2) \textbf{Emilia Dataset}: For comparisons and ablation studies, we used the publicly available Emilia dataset \citep{he2024emilia} (95K hours) including Chinese and English utterances.

\textbf{Architecture Configurations}
We implemented VoxCPM using the Megatron framework, with a 0.5B-parameter configuration, comprising a 24-layer Text-Semantic Language Model (TSLM), initialized from the pre-trained MiniCPM-4-0.5B \citep{team2025minicpm4}\footnote{\url{https://huggingface.co/openbmb/MiniCPM4-0.5B}}, and a randomly initialized 6-layer Residual Acoustic Language Model (RALM). 
The FSQ layer uses 256 dimensions with 9 scalar levels. 
The LocEnc and the LocDiT has 4 Transformers layers, designed for high-efficacy latent extraction and generation.
Detail Configrations are shown in Table \ref{tab: model_config}.

\begin{table}[h]
\centering
\caption{The model architecture of VoxCPM-0.5B.}
\label{tab:voxcpm_arch}
\begin{tabular}{ll}
\toprule
\textbf{Module} & \textbf{Configuration} \\
\midrule
LocEnc & 4 layers, 1024 hidden dim, 4096 FFN dim \\
TSLM   & 24 layers (MiniCPM-4-0.5B initialized), 1024 hidden dim, 4096 FFN dim\\
FSQ & 256 dimensions, 9 quantization levels \\
RALM   & 6 layers, 1024 hidden dim, 4096 FFN dim \\
LocDiT & 4 layers, 1024 hidden dim, 4096 FFN dim\\
Stop Predictor & 3-layer MLP, 1024 hidden dim, 2 output dim\\
patch-size & 2 (that is, TSLM and RALM work in 12.5Hz token rate) \\
AudioVAE &  16kHz waveform $\rightarrow$ 25Hz latents (downsampling at [2, 5, 8, 8]) \\
\bottomrule
\end{tabular}
\label{tab: model_config}
\end{table}

\textbf{Training Details}
We trained two models for comparisons: 
1) \textbf{VoxCPM} was trained with internal large-scale bilingual corpus for 500K iterations using 40 NVIDIA H100 GPUs;
2) \textbf{VoxCPM-Emilia} was trained on the Emilia dataset for 200K iterations using 24 H100 GPUs.
Both VoxCPM and VoxCPM-Emilia used the AdamW optimizer with a peak learning rate of $1\times10^{-4}$ and a Warmup-Stable-Decay (WSD) schedule \citep{hu2024minicpm} which we found essential for optimal convergence. Specifically, the decay phase with annealing to a very low learning rate (combined with batch size doubling) significantly enhances model performance, particularly for zero-shot speaker similarity, as demonstrated in Table \ref{tab:phase_performance}.
All ablation studies followed the same 200K-iteration training protocol on 8 H100 GPUs using the Emilia dataset, employing a fixed learning rate (i.e., without the WSD schedule) of $1\times10^{-4}$
to ensure a consistent comparison.
For LocDiT, we mask the LM condition guidance with a probability ratio of 0.1 for enabling CFG during inference.

\begin{table}[ht]
    \centering
    \caption{Training configurations for VoxCPM variants.}
    \label{tab:training_schedule}
    \begin{tabular}{lccccc}
        \toprule
        \textbf{Model} & \textbf{Phase} & \textbf{Learning Rate} & \textbf{Tokens/Batch} & \textbf{Iterations} & \textbf{GPUs} \\
        \midrule
        VoxCPM & Stable & $1\times10^{-4}$ & 4,096 & 400K & 40 $\times$ H100 \\
        VoxCPM & Decay & $1\times10^{-4} \rightarrow 5\times10^{-6}$ & 8,192 & 100K & 40 $\times$ H100 \\
        \midrule
        VoxCPM-Emilia & Stable & $1\times10^{-4}$ & 4,096 & 150K & 24 $\times$ H100 \\
        VoxCPM-Emilia & Decay & $1\times10^{-4} \rightarrow 5\times10^{-6}$ & 8,192 & 50K & 24 $\times$ H100 \\
        \midrule
        VoxCPM-ablation & Stable & $1\times10^{-4}$ & 4,096 & 200K & 8 $\times$ H100 \\
        \bottomrule
    \end{tabular}
    \label{tab:train_config}
\end{table}

\textbf{Evaluation Metrics and Benchmarks}
We employed comprehensive subjective and objective evaluations. Objective metrics included Word / Character Error Rate (WER / CER) for intelligibility, speaker embedding cosine similarity (SIM) for voice cloning, and DNSMOS for overall quality. 
Subjective evaluation involved Mean Opinion Score (MOS) tests rated by 20 native speakers on naturalness (N-MOS) and speaker similarity (S-MOS) using 5-point scales.
Models were %rigorously 
assessed on two challenging benchmarks: 1) \textbf{SEED-TTS-EVAL}\footnote{\url{https://github.com/BytedanceSpeech/seed-tts-eval}}, focusing on general TTS intelligibility and similarity in English and Chinese, including a ``Hard" set with complex sentences; 2) \textbf{CV3-EVAL}\footnote{\url{https://github.com/FunAudioLLM/CV3-Eval}}, derived from CosyVoice 3 competition, emphasizing  expressive and in-the-wild voice cloning.

\textbf{Baselines}
We compared VoxCPM against a wide range of state-of-the-art open-source TTS systems, including CosyVoice series \citep{du2024cosyvoice1,du2024cosyvoice2}, MaskGCT \citep{wangmaskgct}, F5-TTS \citep{chen2024f5}, SparkTTS \citep{wang2025spark}, FireRedTTS series \citep{guo2024fireredtts, xie2025fireredtts}, IndexTTS 2 \citep{zhou2025indextts2}, HiggsAudio v2 \citep{higgsaudio} and so on.
All baseline results were obtained using official implementations with default settings, or as reported in their original papers.

\subsection{Main Results: Comparison with State-of-the-Art TTS}
As shown in Table \ref{tab:tts_seed_benchmark}, VoxCPM achieves state-of-the-art performance among open-source models on the SEED-TTS-EVAL benchmark. 
It attains an English WER of 1.85\% and a Chinese CER of 0.93\%, surpassing strong competitors like IndexTTS2 and CosyVoice2. 
Concurrently, VoxCPM maintains high speaker similarity, with SIM scores of 72.9\% (EN) and 77.2\% (ZH). 
This demonstrates that the proposed semi-discrete bottleneck effectively balances intelligibility and expressivity by hierarchical semantic-acoustic modeling, mitigating the instability common in continuous models while preserving details often lost in discrete models.
The VoxCPM-Emilia variant, trained on a smaller public dataset, delivers competitive results (EN-WER: 2.34\%, ZH-CER: 1.11\%). 
This highlights the data efficiency and architectural robustness of our approach, as the FSQ bottleneck stabilizes the learning of semantic-acoustic representations even with less training data.
Notably, while DiTAR's phoneme-based approach shows slightly better stability, 
VoxCPM's use of BPE tokens with pre-trained LLM initialization provides superior text understanding capabilities and eliminates dependency on external phonemizers.
Besides, our hierarchical design with residual acoustic modeling reduces the fundamental limitation of direct continuous token modeling, as evidenced in ablation studies.

On the CV3-EVAL benchmark (Table \ref{tab:tts_cv3_combined}), designed to evaluate expressive and in-the-wild performance, 
VoxCPM excels with a ZH-CER of 3.40\% and an EN-WER of 4.04\%. 
Its robustness is further confirmed on the challenging CV3 Hard-Test set, where it achieves an EN-WER of 7.89\%, outperforming even close-sourced CosyVoice 3. 
These results underscore the model's capability to handle complex, realistic inputs, a strength attributed to the RALM's role in recovering fine-grained acoustic details subsequent to the TSLM-FSQ-based semantic-prosodic modeling.

\begin{table}[h!]
    \centering
    \caption{Performance on Seed-TTS-eval Benchmark}
    \label{tab:tts_seed_benchmark}
    \resizebox{\linewidth}{!}{
    \begin{tabular}{lcccccccc}
        \toprule
        \multirow{2}{*}{\textbf{Model}} & \multirow{2}{*}{\textbf{Params}} & \multirow{2}{*}{\textbf{Open-Source}} & \multicolumn{2}{c}{\textbf{EN}} & \multicolumn{2}{c}{\textbf{ZH}} & \multicolumn{2}{c}{\textbf{Hard}}  \\ 
        \cmidrule(lr){4-5} \cmidrule(lr){6-7} \cmidrule(lr){8-9}
        &&& \textbf{WER} $\downarrow$ & \textbf{SIM} $\uparrow$ & \textbf{CER} $\downarrow$ & \textbf{SIM} $\uparrow$ & \textbf{CER} $\downarrow$ & \textbf{SIM} $\uparrow$\\
        \midrule
        \rowcolor{gray!20} MegaTTS3~\citep{jiang2025megatts} & 0.5B & \XSolidBrush & 2.79 & 77.1 & 1.52 & 79.0 & - & - \\
        \rowcolor{gray!20} DiTAR~\citep{jia2025ditar} & 0.6B & \XSolidBrush & 1.69 & 73.5 & 1.02 & 75.3 & - & - \\
        \rowcolor{gray!20} CosyVoice3~\citep{du2025cosyvoice3} & 0.5B & \XSolidBrush & 2.02 & 71.8 & 1.16 & 78.0 & 6.08 & 75.8 \\
        \rowcolor{gray!20} CosyVoice3~\citep{du2025cosyvoice3} & 1.5B & \XSolidBrush & 2.22 & 72.0 & 1.12 & 78.1 & 5.83 & 75.8 \\
        \rowcolor{gray!20} Seed-TTS~\citep{anastassiou2024seed} & - & \XSolidBrush & 2.25 & 76.2 & 1.12 & 79.6 & 7.59 & 77.6 \\
        \rowcolor{gray!20} MiniMax-Speech~\citep{zhang2025minimax} & -& \XSolidBrush & 1.65 & 69.2 & 0.83 & 78.3 & - & - \\
        \midrule
        F5-TTS~\citep{chen2024f5} & 0.3B & \checkmark & 2.00 & 67.0 & 1.53 & 76.0 & 8.67 & 71.3 \\
        MaskGCT~\citep{wangmaskgct} & & \checkmark & 2.62 & \underline{71.7} & 2.27 & \textbf{77.4} & - & - \\
        CosyVoice~\citep{du2024cosyvoice1} & 0.3B & \checkmark & 4.29 & 60.9 & 3.63 & 72.3 & 11.75 & 70.9 \\
        CosyVoice2~\citep{du2024cosyvoice2} & 0.5B & \checkmark & 3.09 & 65.9 & 1.38 & 75.7 & \textbf{6.83} & 72.4 \\
        SparkTTS~\citep{wang2025spark} & 0.5B & \checkmark & 3.14 & 57.3 & 1.54 & 66.0 & - & - \\
        FireRedTTS~\citep{guo2024fireredtts} & 0.5B & \checkmark & 3.82 & 46.0 & 1.51 & 63.5 & 17.45 & 62.1 \\
        FireRedTTS-2~\citep{xie2025fireredtts} &  & \checkmark & 1.95 & 66.5 & 1.14 & 73.6 & - & - \\
        Qwen2.5-Omni~\citep{xu2025qwen2} & 7B & \checkmark & 2.72 & 63.2 & 1.70 & 75.2 & 7.97 & \underline{74.7} \\
        OpenAudio-s1-mini~\citep{openaudios1} & 0.5B & \checkmark & \underline{1.94} & 55.0 & 1.18 & 68.5 & 23.37 & 64.3 \\
        IndexTTS 2~\citep{zhou2025indextts2} & 1.5B & \checkmark & 2.23 & 70.6 & \underline{1.03} & 76.5 & \underline{7.12} & \textbf{75.5} \\
        VibeVoice~\citep{peng2025vibevoice} & 1.5B & \checkmark & 3.04 & 68.9 & 1.16 & 74.4 & - & - \\
        HiggsAudio-v2~\citep{higgsaudio} & 3B & \checkmark & 2.44 & 67.7 & 1.50 & 74.0 & 55.07 & 65.6 \\
        \midrule
\textbf{VoxCPM-Emilia} & 0.5B & \checkmark & 2.34 & 68.1 & 1.11 & 74.0 & 12.46 & 69.8 \\
 \textbf{VoxCPM} & 0.5B & \checkmark & \textbf{1.85} & \textbf{72.9} & \textbf{0.93} & \underline{77.2} & 8.87 & 73.0 \\
        \bottomrule
    \end{tabular}
    }
\end{table}

\begin{table}[ht]
    \centering
    \caption{Performance on CV3-eval Benchmark. *denotes close-sourced systems.}
    \label{tab:tts_cv3_combined}
    \resizebox{1.0\linewidth}{!}{
    \begin{tabular}{lcccccccc}
        \toprule
        \multirow{2}{*}{\textbf{Model}} & \multicolumn{2}{c}{\textbf{CV3-EVAL}} & \multicolumn{3}{c}{\textbf{CV3-Hard-ZH}} & \multicolumn{3}{c}{\textbf{CV3-Hard-EN}} \\
        \cmidrule(lr){2-3} \cmidrule(lr){4-6} \cmidrule(lr){7-9}
        & \textbf{ZH-CER} $\downarrow$ & \textbf{EN-WER} $\downarrow$ & \textbf{CER} $\downarrow$ & \textbf{SIM} $\uparrow$ & \textbf{DNSMOS}$\uparrow$  & \textbf{WER} $\downarrow$ & \textbf{SIM} $\uparrow$ & \textbf{DNSMOS}$\uparrow$  \\
        \midrule
        F5-TTS & 5.47 & 8.90 & - & - & - & - & - & - \\
        SparkTTS & 5.15 & 11.0 & - & - & - & - & - & - \\
        GPT-Sovits & 7.34 & 12.5 & - & - & - & - & - & - \\
        CosyVoice2 & 4.08 & 6.32 & 12.58 & 72.6 & \textbf{3.81} & 11.96 & 66.7 & \textbf{3.95} \\
        OpenAudio-s1-mini & 4.00 & 5.54 & 18.1 & 58.2 & 3.77 & 12.4 & 55.7 & 3.89 \\
        IndexTTS2 & 3.58 & 4.45 & 12.8 & \textbf{74.6} & 3.65 & 8.78 & \textbf{74.5} & 3.80 \\
        HiggsAudio-v2 & 9.54 & 7.89 & 41.0 & 60.2 & 3.39 & 10.3 & 61.8 & 3.68 \\
        \rowcolor{gray!20} CosyVoice3-0.5B* & 3.89 & 5.24 & 14.15 & 78.6 & 3.75 & 9.04 & 75.9 & 3.92 \\
        \rowcolor{gray!20} CosyVoice3-1.5B* & 3.91 & 4.99 & 9.77 & 78.5 & 3.79 & 10.55 & 76.1 & 3.95 \\
        \midrule
        \textbf{VoxCPM-Emilia} & 4.47 & 5.23 & 22.2 & 62.6 & 3.47 & 10.00 & 62.6 & 3.68 \\
        \textbf{VoxCPM} & \textbf{3.40} & \textbf{4.04} & \textbf{12.9} & 66.1 & 3.59 & \textbf{7.89} & 64.3 & 3.74 \\
        \bottomrule
    \end{tabular}
    }
\end{table}

Subjective evaluations (Table \ref{tab:tts_mos}) further validate the objective findings, with VoxCPM achieving competitive performance across both languages. On English tests, VoxCPM obtains the highest scores in speaker similarity and good results in naturalness.
For Chinese, while VoxCPM trails IndexTTS 2 in naturalness, it achieves slightly superior speaker similarity. This pattern suggests that VoxCPM excels at voice cloning consistency, while IndexTTS 2 may have advantages in prosodic naturalness for Chinese.
VoxCPM-Emilia shows competitive speaker similarity but relatively lower naturalness, highlighting the impact of training data scale. 

\begin{table}[h!]
    \centering
    \caption{Subjective Evaluations in terms of Naturalness and Speaker Similarity.}
    \label{tab:tts_mos}
    \resizebox{0.75\linewidth}{!}{
    \begin{tabular}{lcccc}
    \toprule
    \multirow{2}{*}{\textbf{Model}} & \multicolumn{2}{c}{\textbf{ZH}} & \multicolumn{2}{c}{\textbf{EN}} \\
    \cmidrule(lr){2-3} \cmidrule(lr){4-5}
    & \textbf{N-MOS} & \textbf{S-MOS} & \textbf{N-MOS} & \textbf{S-MOS} \\
    \midrule
    MaskGCT & $3.20\pm0.11$ & $3.77\pm0.11$  & $3.84\pm0.11$ & $4.00\pm0.10$ \\
    CosyVoice 2  & $3.38\pm0.12$ & $4.01\pm0.10$ & $\mathbf{4.14\pm0.09}$ & $3.97\pm0.10$ \\
    IndexTTS 2 & $\mathbf{4.25\pm0.09}$ & \underline{$4.05\pm0.09$} & $4.03\pm0.10$ & \underline{$4.16\pm0.09$} \\
    \textbf{VoxCPM-Emilia} & $3.79\pm0.12$ & $3.99\pm0.11$ & $3.91\pm0.10$ & $4.10\pm0.09$ \\
    \textbf{VoxCPM} & \underline{$4.10\pm0.10$} & $\mathbf{4.11\pm0.10}$ & \underline{$4.11\pm0.09$} & $\mathbf{4.18\pm0.09}$ \\
    \bottomrule
    \end{tabular}
    }
\end{table}

\subsection{Ablation Study: Effect of the Semi-discrete Bottleneck}
As shown in Table \ref{tab:fsq_dimension}, the ablation studies on the FSQ bottleneck dimensionality provide critical insights. 
The catastrophic performance degradation of the purely continuous model (w/o FSQ), especially on hard cases (ZH-CER: 24.92\%), validates our core hypothesis: entangling semantic planning and acoustic rendering in a continuous space leads to instability. Without the inductive bias imposed by FSQ, the model struggles to separate these tasks even with a hierarchical design, resulting in error accumulation on complex utterances.

\begin{table}[h!]
\centering
\caption{FSQ dimension selection study on the Emilia dataset. \textit{Note:} The 256-dim was selected for the final VoxCPM configuration, with the understanding that larger training datasets needs more powerful modeling capabilities.}
\begin{tabular}{lcccccc}
\toprule
\multirow{2}{*}{\textbf{Model Setting}} & \multicolumn{2}{c}{\textbf{EN}} & \multicolumn{2}{c}{\textbf{ZH}} & \multicolumn{2}{c}{\textbf{ZH-hard case}} \\
\cmidrule(lr){2-3} \cmidrule(lr){4-5} \cmidrule(lr){6-7}
 & \textbf{WER} $\downarrow$ & \textbf{SIM} $\uparrow$ & \textbf{CER} $\downarrow$ & \textbf{SIM} $\uparrow$ & \textbf{CER} $\downarrow$ & \textbf{SIM} $\uparrow$ \\
\midrule
w FSQ: d4s9 & 5.18  & 59.3 & 4.05 & 68.0 & 19.55 & 62.3 \\
w FSQ: d16s9 & 3.22 & 60.4 & 1.87 & \underline{70.5} & \textbf{14.42} & \textbf{66.2} \\
w FSQ: d64s9 & 3.22 & 61.1 & 2.14 & 69.8 & 17.48 & 65.1 \\
w FSQ: d128s9 & 3.43 & \underline{62.2} & \textbf{1.67} & \textbf{70.7} & \underline{16.76} & \underline{65.7} \\
w FSQ: d256s9 & \textbf{2.98} & \textbf{62.6} & \underline{1.77} & 70.4 & 18.19 & 64.9 \\
w FSQ: d1024s9 & \underline{3.07} & 62.0 & 2.38 & 69.8 & 20.38 & 64.7 \\
w/o FSQ: d1024s$\infty$ & 3.67 & 62.1 & 2.30 & 69.6 & 24.92 & 63.5 \\
\bottomrule
\end{tabular}
\label{tab:fsq_dimension}
\end{table}

The optimal performance observed at %intermediate 
FSQ levels (FSQ-d128/d256) reveals a key trade-off. Lower dimensions (e.g., FSQ-d4) over-constrain the representation, limiting prosodic capacity. Higher dimensions (e.g., FSQ-d1024) provide insufficient discretization strength, allowing task entanglement to persist. The peak at FSQ-d256 indicates the bottleneck creates an effective ``summary space": discrete enough to stabilize long-range semantic planning yet continuous enough to retain crucial prosodic and speaker information, thereby enforcing a beneficial division of labor within the model.

\subsection{Ablation Study: Effect of Residual Acoustic Modeling}

As shown in Table \ref{tab:ralm_ablation}, the ablation studies about the residual language modeling validate our core architectural innovations.
Notably, the purely continuous variant (w/o RALM: TSLM $\rightarrow$ LocDiT) —analogous to DiTAR's approach—shows significantly degraded performance, particularly on challenging cases.
The performance gap persists across different TSLM configurations, confirming that the challenge is fundamental to the learning objective rather than parameter allocation.
This conclusively demonstrates the advantage of our explicit separation between semantic and acoustic modeling.
Secondly, the critical role of residual acoustic input is further evidenced by the substantial degradation when ablating original acoustic embeddings (w/o $E_{<i}$ in RALM), highlighting that the RALM requires fine-grained acoustic information to accurately recover acoustic details.
Finally, the best performance of the default setting demonstrates the effectiveness of the residual connection. 
By summing the TSLM and RALM hidden states, the model explicitly delegates semantic-prosodic planning to the TSLM and acoustic refinement to the RALM, achieving optimal integration.

\begin{table}[h]
\centering
\caption{Ablation Studies about core architecture designs.}
\resizebox{1.0\linewidth}{!}{
\begin{tabular}{lcccccc}
\toprule
\multirow{2}{*}{\textbf{Model Setting}} & \multicolumn{2}{c}{\textbf{EN}} & \multicolumn{2}{c}{\textbf{ZH}} & \multicolumn{2}{c}{\textbf{ZH-hard case}} \\
\cmidrule(lr){2-3} \cmidrule(lr){4-5} \cmidrule(lr){6-7} 
 & \textbf{WER} $\downarrow$ & \textbf{SIM} $\uparrow$ & \textbf{CER} $\downarrow$ & \textbf{SIM} $\uparrow$ & \textbf{CER} $\downarrow$ & \textbf{SIM} $\uparrow$ \\
 \midrule
\textbf{default setting} & 2.98 & 62.6 & 1.77 & 70.4 & 18.19 & 64.9 \\
w/o RALM: TSLM (24 layers) $\rightarrow$ \text{LocDiT} & 4.34 & 61.8 & 3.05 & 69.4 & 25.00 & 63.8 \\
w/o RALM: TSLM (30 layers) $\rightarrow$ \text{LocDiT} & 5.35 & 62.6 & 3.46 & 69.8 & 30.40 & 63.9 \\
w/o $E_{<i}$ in RALM: TSLM $\rightarrow$ \text{ALM} $\rightarrow$ \text{LocDiT} & 4.91 & 60.9 & 4.94 & 68.1 & 27.17 & 61.7 \\
w/o $h^{\text{residual}}$ in condition: TSLM $\rightarrow$ \text{FSQ} $\rightarrow$ \text{LocDiT} & 3.86 & 58.3 & 3.05 & 67.6 & 23.65 & 61.7 \\
\bottomrule
\end{tabular}
}
\label{tab:ralm_ablation}
\end{table}

\subsection{Effect of Training Phase on Performance}
As mentioned in Table \ref{tab:train_config}, the two-phase Warmup-Stable-Decay (WSD) learning rate schedule is critical for achieving optimal model performance. The initial Stable phase allows the model to converge reliably to a strong baseline. The subsequent Decay phase is then essential for refining the model, particularly for improving its zero-shot voice similarity capabilities.

The performance gains from this two-phase strategy are substantiated in Table \ref{tab:phase_performance}. 
Compared to the Stable phase, the Decay phase achieves consistent improvements across all metrics: reducing word error rates, while simultaneously enhancing speaker similarity. Most notably, the model demonstrates a remarkable leap in robustness on challenging cases, with the CER on ZH-Hard dropping from 13.22\% to 8.87\%, alongside a 4.4-point SIM improvement.

\begin{table}[ht]
    \centering
    \caption{Performance across training phases.}
    \label{tab:phase_performance}
    \begin{tabular}{lcccccc}
        \toprule
        \multirow{2}{*}{\textbf{Phase}} & \multicolumn{2}{c}{\textbf{EN}} & \multicolumn{2}{c}{\textbf{ZH}} & \multicolumn{2}{c}{\textbf{ZH-Hard Case}} \\
        \cmidrule(lr){2-3} \cmidrule(lr){4-5} \cmidrule(lr){6-7}
        & \textbf{WER} $\downarrow$ & \textbf{SIM} $\uparrow$ & \textbf{CER} $\downarrow$ & \textbf{SIM} $\uparrow$ & \textbf{CER} $\downarrow$ & \textbf{SIM} $\uparrow$ \\
        \midrule
        Stable& 2.05 & 69.7 & 0.99 & 75.1 & 13.22 & 68.6 \\
        Decay & 1.85 & 72.9 & 0.93 & 77.2 & 8.87 & 73.0 \\
        \bottomrule
    \end{tabular}
\end{table}

\subsection{Effect of LM Guidance on LocDiT}
To investigate the influence of Classifier-Free Guidance (CFG) and identify the optimal inference setting, we tested different CFG value, that is, the LM (the sum of TSLM-FSQ hidden and RALM hidden) guidance on LocDiT.
As detailed in Table \ref{tab:cfg}, the CFG scale exerts a profound and non-monotonic influence on the trade-off between speech intelligibility and speaker similarity. 
The absence of CFG (a scale of 1.0) results in poor performance, characterized by high error rates and low similarity scores, as the model lacks sufficient incentive to strongly condition on the linguistic input. 
Employing a moderate CFG value of 2.0 yields the optimal balance, effectively enhancing voice similarity without compromising intelligibility, while higher values ($\geq$3.0) degraded intelligibility significantly.

\begin{table}[h]
\centering
\small
 \caption{Effect of LM guidance on LocDiT, tested with \textbf{VoxCPM}.}
\begin{tabular}{lcccccc}
\toprule
\multirow{2}{*}{\textbf{CFG Value}} & \multicolumn{2}{c}{\textbf{EN}} & \multicolumn{2}{c}{\textbf{ZH}} & \multicolumn{2}{c}{\textbf{ZH-hard case}} \\
\cmidrule(lr){2-3} \cmidrule(lr){4-5} \cmidrule(lr){6-7}
 & \textbf{WER} $\downarrow$ & \textbf{SIM} $\uparrow$ & \textbf{CER} $\downarrow$ & \textbf{SIM} $\uparrow$ & \textbf{CER} $\downarrow$ & \textbf{SIM} $\uparrow$ \\
\midrule
1.0 (w/o CFG) & 16.32 & 55.1 & 14.47 & 61.5 & 56.87 & 43.0\\
1.5 & 1.86 & 72.1 & 1.16 & 77.0 & 9.60 & 73.9 \\
2.0 & 1.85 & 72.9 & 0.93 & 77.2 & 8.87 & 73.0 \\
3.0 & 2.16 & 71.4 & 1.12 & 74.7 & 13.22 & 65.0 \\ 
5.0  & 12.78 & 60.7 & 17.23 & 59.4 & 48.46& 39.9 \\ 
\bottomrule
\end{tabular}
\label{tab:cfg}
\end{table}

\subsection{Analysis and Discussion}

\textbf{Visual Analysis of Hierarchical Representations} 
\begin{figure}[h]
    \centering
    \includegraphics[width=1.0\textwidth]{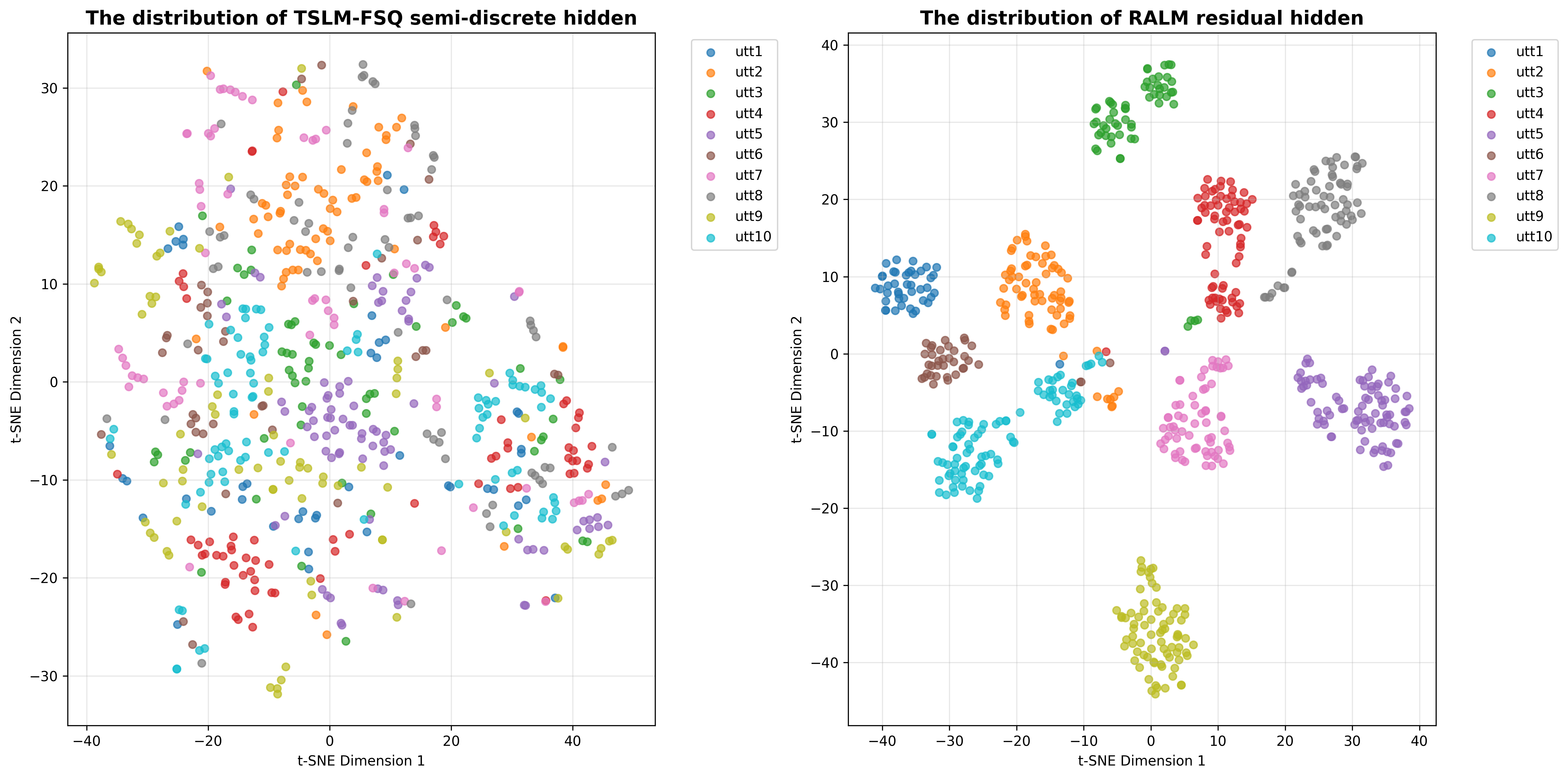}
    \caption{The T-SNE visualization of latent space distributions in zero-shot voice cloning task.}
    \label{fig:tsne1}
\end{figure}

\begin{figure}[h]
    \centering
    \includegraphics[width=1.0\textwidth]{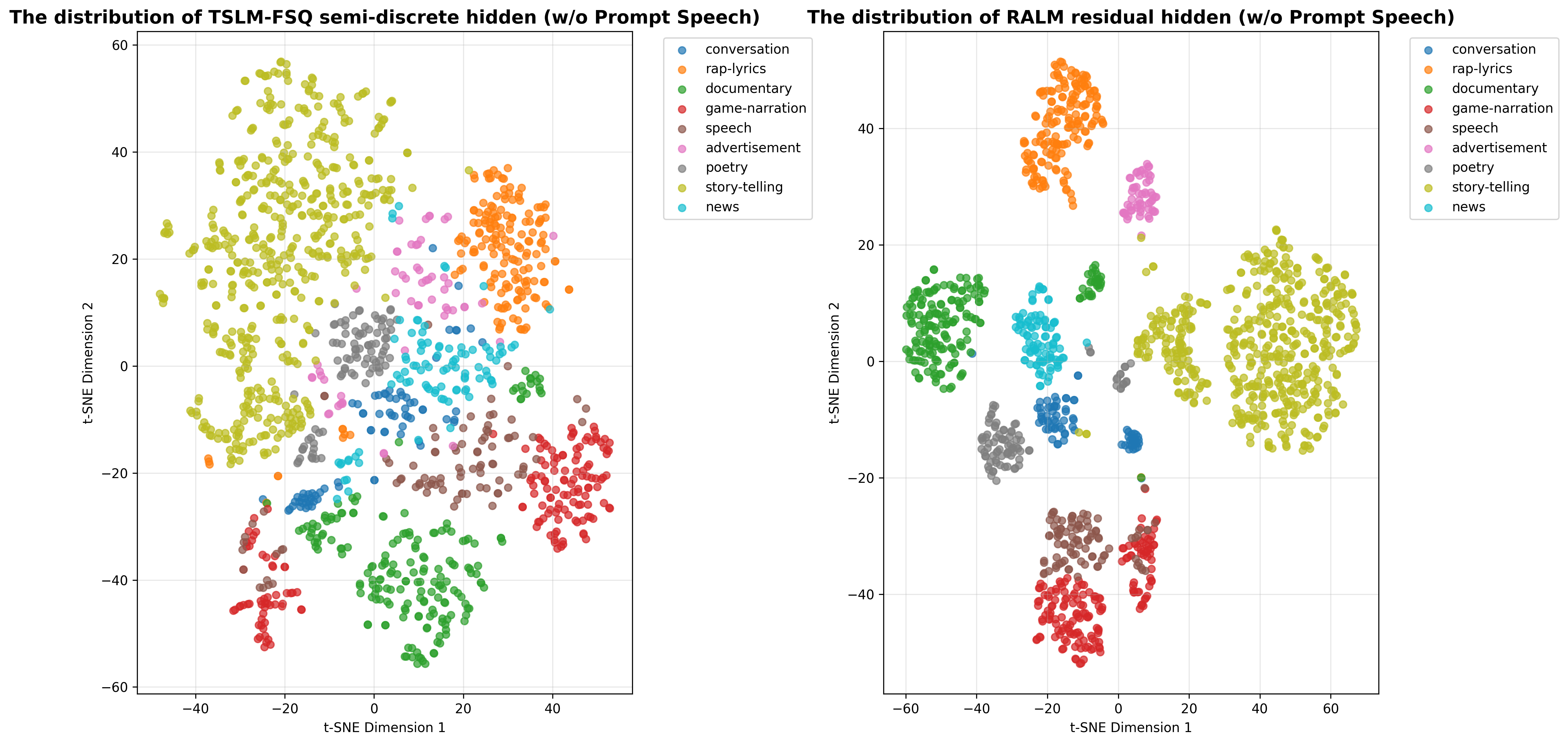}
    \caption{The T-SNE visualization of latent space distributions in text-to-speech task, without prompt speech.}
    \label{fig:tsne2}
\end{figure}

To validate our core hypothesis of learned implicit semantic-acoustic disentanglement, we conducted a t-SNE visualization of the internal representations in our hierarchical model.
The resulting distributions, shown in Figures \ref{fig:tsne1} and \ref{fig:tsne2}, empirically confirm the specialized roles of the TSLM and the RALM.
Figure \ref{fig:tsne1} illustrates the model's behavior in a zero-shot voice cloning task, where each color corresponds to a distinct utterance from an unseen speaker.
The TSLM-FSQ outputs form semantic-prosodic structure closely tied to text content, while the RALM residuals  exhibit strong speaker-related variations for acoustic rendering, confirming their specialized roles in content planning and acoustic refinement.
Figure \ref{fig:tsne2} further demonstrates the VoxCPM's capability to infer appropriate prosody and style directly from text, when not using any speech prompt. 
When processing different text genres (news, poetry, conversation), TSLM-FSQ representations cluster by semantic category, showing that the pre-trained language model backbone effectively infers appropriate prosodic patterns directly from text content. 
For example, embeddings for ``news" group together, separate from ``story-telling" or ``rap-lyrics."
The RALM outputs display greater within-category variation, indicating its role in adding fine-grained acoustic nuances to the semantic-prosodic plan.

\textbf{Expressive and Context-Aware Synthesis Capabilities}
Beyond quantitative metrics, VoxCPM shows
good expressive and context-aware synthesis capabilities directly from text benfiting from the architecture design and training data. 
The powerful pre-trained LM backbone provides inherent text understanding, enabling appropriate prosodic variations across different content types, as mentioned above. 
When not using prompt speech, the model tends to express suitable style from contextual cues, also shown in Figure \ref{fig:tsne2}. 
We strongly recommend readers to listen our demo samples\footnote{\url{https://openbmb.github.io/VoxCPM-demopage/}}.

\textbf{Scalability and Efficiency} 
The performance improvement from VoxCPM-Emilia to VoxCPM highlights the architecture's scalability with increased data. 
The hierarchical design allows larger models to effectively utilize increased capacity for learning complex patterns. 
In terms of inference efficiency, VoxCPM-0.5B achieves a real-time factor (RTF) of 0.17 on a single NVIDIA RTX 4090, confirming practical deployment feasibility.

\section{Conclusion}
In this work, we present a novel tokenizer free TTS model VoxCPM to achieve context-aware speech generation and true-to-life voice cloning.
It resolves the fundamental trade-off between expressivity and stability in text-to-speech synthesis by introducing a unified, end-to-end framework based on hierarchical semantic-acoustic modeling with semi-discrete residual representations. 
Our approach leverages a differentiable quantization bottleneck to induce a natural separation of concerns: a text-semantic language model captures high-level semantic-prosodic structure, while a residual acoustic model recovers fine-grained details. 
This eliminates the dependency on external speech tokenizers and mitigates the error accumulation that plagues purely continuous autoregressive models. 
Extensive experiments demonstrate that our model achieves state-of-the-art zero-shot TTS performance among open-source systems, excelling in both intelligibility and speaker similarity. The success of VoxCPM validates that learning structured, regularized latent spaces provides a principled foundation for expressive generative audio modeling.
 
\textbf{Limitations}
Despite these advancements, our work still has several limitations. 
First, the model's multilingual capability remains limited, as it is primarily optimized for Chinese and English, with uncertain generalization to other languages. 
Second, the controllability of speech attributes—such as fine-grained prosody and emotional expression—is still constrained, lacking both intuitive user guidance and precise adjustment mechanisms. 
Finally, the current AudioVAE only supports 16kHz audio generation, which restricts perceptual quality and falls short of high-fidelity application requirements that typically demand 24kHz or 44.1kHz sampling rates. 
These limitations point to meaningful directions for future research.

\textbf{Ethics statement}
Since our zero-shot TTS model achieves high-quality speech synthesis with the ability to closely mimic speaker characteristics, it carries potential risks of misuse. These risks include, but are not limited to, spoofing voice authentication systems or impersonating a specific speaker without their consent. Our experiments were conducted under the assumption that the use of any reference speaker's voice is authorized and intended for legitimate synthesis purposes. To mitigate these risks, we strongly advocate for the development of robust synthesized speech detection algorithms. Furthermore, we believe it is crucial to establish clear ethical guidelines and reporting mechanisms for the responsible deployment of such technology.

\section{Contributors}
\modelname{} is a collaborative release by the Tsinghua Shenzhen International Graduate School (SIGS) Human-Computer Speech Interaction Lab (THUHCSI), Natural Language Processing Lab at Tsinghua University (THUNLP) and ModelBest. We would also like to thank the OpenBMB community for all their support.

\textbf{Core Contributors} \quad
Yixuan Zhou, Guoyang Zeng, Xin Liu, Xiang Li, Renjie Yu, Ziyang Wang, Runchuan Ye, Weiyue Sun, Jiancheng Gui, Kehan Li, Zhiyong Wu, Zhiyuan Liu

\textbf{Other Contributors (Alphabetical order)} \quad
Biyuan Lin, Chao Jia, Chenzhe Jing, Hongyu Liu, Jie Cai, Jie Zhou, Junshao Guo, Lei Chen, Rongting Tang, Rui Li, Ruiqi Shao, Qundong Shi, Shuo Wang, Siyuan Huang, Shun Lei, Wenxi Yang, Xiaoshuang Wang, Yihang He, Zichao Nie

\newpage

\bibliographystyle{citation}
\bibliography{citation}

\begin{thebibliography}{50}
\providecommand{\natexlab}[1]{#1}
\providecommand{\url}[1]{\texttt{#1}}
\expandafter\ifx\csname urlstyle\endcsname\relax
  \providecommand{\doi}[1]{doi: #1}\else
  \providecommand{\doi}{doi: \begingroup \urlstyle{rm}\Url}\fi

\bibitem[An et~al.(2025)An, Zhang, Gao, Li, Peng, Wang, Du, Zhao, Gao, and Li]{an2025mela}
Keyu An, Zhiyu Zhang, Changfeng Gao, Yabin Li, Zhendong Peng, Haoxu Wang, Zhihao Du, Han Zhao, Zhifu Gao, and Xiangang Li.
\newblock Mela-tts: Joint transformer-diffusion model with representation alignment for speech synthesis.
\newblock \emph{arXiv preprint arXiv:2509.14784}, 2025.

\bibitem[Anastassiou et~al.(2024)Anastassiou, Chen, Chen, Chen, Chen, Chen, Cong, Deng, Ding, Gao, et~al.]{anastassiou2024seed}
Philip Anastassiou, Jiawei Chen, Jitong Chen, Yuanzhe Chen, Zhuo Chen, Ziyi Chen, Jian Cong, Lelai Deng, Chuang Ding, Lu~Gao, et~al.
\newblock Seed-tts: A family of high-quality versatile speech generation models.
\newblock \emph{arXiv preprint arXiv:2406.02430}, 2024.

\bibitem[Baas et~al.(2025)Baas, Scholtz, Mehta, Dyson, Prakash, and Kamper]{baas2025mars6}
Matthew Baas, Pieter Scholtz, Arnav Mehta, Elliott Dyson, Akshat Prakash, and Herman Kamper.
\newblock Mars6: A small and robust hierarchical-codec text-to-speech model.
\newblock In \emph{ICASSP 2025-2025 IEEE International Conference on Acoustics, Speech and Signal Processing (ICASSP)}, pp.\  1--5. IEEE, 2025.

\bibitem[Borsos et~al.(2023{\natexlab{a}})Borsos, Marinier, Vincent, Kharitonov, Pietquin, Sharifi, Roblek, Teboul, Grangier, Tagliasacchi, et~al.]{borsos2023audiolm}
Zal{\'a}n Borsos, Rapha{\"e}l Marinier, Damien Vincent, Eugene Kharitonov, Olivier Pietquin, Matt Sharifi, Dominik Roblek, Olivier Teboul, David Grangier, Marco Tagliasacchi, et~al.
\newblock Audiolm: a language modeling approach to audio generation.
\newblock \emph{IEEE/ACM transactions on audio, speech, and language processing}, 31:\penalty0 2523--2533, 2023{\natexlab{a}}.

\bibitem[Borsos et~al.(2023{\natexlab{b}})Borsos, Sharifi, Vincent, Kharitonov, Zeghidour, and Tagliasacchi]{borsos2023soundstorm}
Zal{\'a}n Borsos, Matt Sharifi, Damien Vincent, Eugene Kharitonov, Neil Zeghidour, and Marco Tagliasacchi.
\newblock Soundstorm: Efficient parallel audio generation.
\newblock \emph{arXiv preprint arXiv:2305.09636}, 2023{\natexlab{b}}.

\bibitem[BosonAI(2025)]{higgsaudio}
BosonAI.
\newblock Higgs audio v2: Redefining expressiveness in audio generation.
\newblock \emph{https://github.com/boson-ai/higgs-audio}, 2025.

\bibitem[Casanova et~al.(2024)Casanova, Davis, G{\"o}lge, G{\"o}knar, Gulea, Hart, Aljafari, Meyer, Morais, Olayemi, et~al.]{casanova2024xtts}
Edresson Casanova, Kelly Davis, Eren G{\"o}lge, G{\"o}rkem G{\"o}knar, Iulian Gulea, Logan Hart, Aya Aljafari, Joshua Meyer, Reuben Morais, Samuel Olayemi, et~al.
\newblock Xtts: a massively multilingual zero-shot text-to-speech model.
\newblock \emph{arXiv preprint arXiv:2406.04904}, 2024.

\bibitem[Chen et~al.(2025)Chen, Wang, Wu, Zhang, Zhou, Liu, Chen, Liu, Wang, Li, et~al.]{chen2025neural}
Sanyuan Chen, Chengyi Wang, Yu~Wu, Ziqiang Zhang, Long Zhou, Shujie Liu, Zhuo Chen, Yanqing Liu, Huaming Wang, Jinyu Li, et~al.
\newblock Neural codec language models are zero-shot text to speech synthesizers.
\newblock \emph{IEEE Transactions on Audio, Speech and Language Processing}, 2025.

\bibitem[Chen et~al.(2024)Chen, Niu, Ma, Deng, Wang, Zhao, Yu, and Chen]{chen2024f5}
Yushen Chen, Zhikang Niu, Ziyang Ma, Keqi Deng, Chunhui Wang, Jian Zhao, Kai Yu, and Xie Chen.
\newblock F5-tts: A fairytaler that fakes fluent and faithful speech with flow matching.
\newblock \emph{CoRR}, 2024.

\bibitem[D{\'e}fossez et~al.(2022)D{\'e}fossez, Copet, Synnaeve, and Adi]{defossez2022high}
Alexandre D{\'e}fossez, Jade Copet, Gabriel Synnaeve, and Yossi Adi.
\newblock High fidelity neural audio compression.
\newblock \emph{arXiv preprint arXiv:2210.13438}, 2022.

\bibitem[Deng et~al.(2025)Deng, Zhou, Shu, Wang, and Wang]{deng2025indextts}
Wei Deng, Siyi Zhou, Jingchen Shu, Jinchao Wang, and Lu~Wang.
\newblock Indextts: An industrial-level controllable and efficient zero-shot text-to-speech system.
\newblock \emph{arXiv preprint arXiv:2502.05512}, 2025.

\bibitem[Du et~al.(2024{\natexlab{a}})Du, Chen, Zhang, Hu, Lu, Yang, Hu, Zheng, Gu, Ma, et~al.]{du2024cosyvoice1}
Zhihao Du, Qian Chen, Shiliang Zhang, Kai Hu, Heng Lu, Yexin Yang, Hangrui Hu, Siqi Zheng, Yue Gu, Ziyang Ma, et~al.
\newblock Cosyvoice: A scalable multilingual zero-shot text-to-speech synthesizer based on supervised semantic tokens.
\newblock \emph{arXiv preprint arXiv:2407.05407}, 2024{\natexlab{a}}.

\bibitem[Du et~al.(2024{\natexlab{b}})Du, Wang, Chen, Shi, Lv, Zhao, Gao, Yang, Gao, Wang, et~al.]{du2024cosyvoice2}
Zhihao Du, Yuxuan Wang, Qian Chen, Xian Shi, Xiang Lv, Tianyu Zhao, Zhifu Gao, Yexin Yang, Changfeng Gao, Hui Wang, et~al.
\newblock Cosyvoice 2: Scalable streaming speech synthesis with large language models.
\newblock \emph{arXiv preprint arXiv:2412.10117}, 2024{\natexlab{b}}.

\bibitem[Du et~al.(2025)Du, Gao, Wang, Yu, Zhao, Wang, Lv, Wang, Ni, Shi, et~al.]{du2025cosyvoice3}
Zhihao Du, Changfeng Gao, Yuxuan Wang, Fan Yu, Tianyu Zhao, Hao Wang, Xiang Lv, Hui Wang, Chongjia Ni, Xian Shi, et~al.
\newblock Cosyvoice 3: Towards in-the-wild speech generation via scaling-up and post-training.
\newblock \emph{arXiv preprint arXiv:2505.17589}, 2025.

\bibitem[Guo et~al.(2024)Guo, Hu, Liu, Shen, Tang, Wu, Xie, Xie, and Xu]{guo2024fireredtts}
Hao-Han Guo, Yao Hu, Kun Liu, Fei-Yu Shen, Xu~Tang, Yi-Chen Wu, Feng-Long Xie, Kun Xie, and Kai-Tuo Xu.
\newblock Fireredtts: A foundation text-to-speech framework for industry-level generative speech applications.
\newblock \emph{arXiv preprint arXiv:2409.03283}, 2024.

\bibitem[Han et~al.(2025)Han, Hao, Chen, Xiong, He, Zhang, Cao, Liu, Li, Zhang, et~al.]{han2025quantize}
Yichen Han, Xiaoyang Hao, Keming Chen, Weibo Xiong, Jun He, Ruonan Zhang, Junjie Cao, Yue Liu, Bowen Li, Dongrui Zhang, et~al.
\newblock Quantize more, lose less: Autoregressive generation from residually quantized speech representations.
\newblock \emph{arXiv preprint arXiv:2507.12197}, 2025.

\bibitem[He et~al.(2024)He, Shang, Wang, Li, Gu, Hua, Liu, Yang, Li, Shi, et~al.]{he2024emilia}
Haorui He, Zengqiang Shang, Chaoren Wang, Xuyuan Li, Yicheng Gu, Hua Hua, Liwei Liu, Chen Yang, Jiaqi Li, Peiyang Shi, et~al.
\newblock Emilia: An extensive, multilingual, and diverse speech dataset for large-scale speech generation.
\newblock In \emph{2024 IEEE Spoken Language Technology Workshop (SLT)}, pp.\  885--890. IEEE, 2024.

\bibitem[Hu et~al.(2024)Hu, Tu, Han, He, Cui, Long, Zheng, Fang, Huang, Zhao, et~al.]{hu2024minicpm}
Shengding Hu, Yuge Tu, Xu~Han, Chaoqun He, Ganqu Cui, Xiang Long, Zhi Zheng, Yewei Fang, Yuxiang Huang, Weilin Zhao, et~al.
\newblock Minicpm: Unveiling the potential of small language models with scalable training strategies.
\newblock \emph{arXiv preprint arXiv:2404.06395}, 2024.

\bibitem[Jia et~al.(2025)Jia, Chen, Chen, Du, Wu, Cong, Zhuang, Li, Wei, Wang, et~al.]{jia2025ditar}
Dongya Jia, Zhuo Chen, Jiawei Chen, Chenpeng Du, Jian Wu, Jian Cong, Xiaobin Zhuang, Chumin Li, Zhen Wei, Yuping Wang, et~al.
\newblock Ditar: Diffusion transformer autoregressive modeling for speech generation.
\newblock \emph{arXiv preprint arXiv:2502.03930}, 2025.

\bibitem[Jiang et~al.(2025)Jiang, Ren, Li, Ji, Zhang, Ye, Zhang, Jionghao, Yang, Zuo, et~al.]{jiang2025megatts}
Ziyue Jiang, Yi~Ren, Ruiqi Li, Shengpeng Ji, Boyang Zhang, Zhenhui Ye, Chen Zhang, Bai Jionghao, Xiaoda Yang, Jialong Zuo, et~al.
\newblock Megatts 3: Sparse alignment enhanced latent diffusion transformer for zero-shot speech synthesis.
\newblock \emph{arXiv preprint arXiv:2502.18924}, 2025.

\bibitem[Kharitonov et~al.(2023)Kharitonov, Vincent, Borsos, Marinier, Girgin, Pietquin, Sharifi, Tagliasacchi, and Zeghidour]{kharitonov2023speak}
Eugene Kharitonov, Damien Vincent, Zal{\'a}n Borsos, Rapha{\"e}l Marinier, Sertan Girgin, Olivier Pietquin, Matt Sharifi, Marco Tagliasacchi, and Neil Zeghidour.
\newblock Speak, read and prompt: High-fidelity text-to-speech with minimal supervision.
\newblock \emph{Transactions of the Association for Computational Linguistics}, 11:\penalty0 1703--1718, 2023.

\bibitem[Kumar et~al.(2023)Kumar, Seetharaman, Luebs, Kumar, and Kumar]{kumar2023high}
Rithesh Kumar, Prem Seetharaman, Alejandro Luebs, Ishaan Kumar, and Kundan Kumar.
\newblock High-fidelity audio compression with improved rvqgan.
\newblock \emph{Advances in Neural Information Processing Systems}, 36:\penalty0 27980--27993, 2023.

\bibitem[Le et~al.(2023)Le, Vyas, Shi, Karrer, Sari, Moritz, Williamson, Manohar, Adi, Mahadeokar, et~al.]{le2023voicebox}
Matthew Le, Apoorv Vyas, Bowen Shi, Brian Karrer, Leda Sari, Rashel Moritz, Mary Williamson, Vimal Manohar, Yossi Adi, Jay Mahadeokar, et~al.
\newblock Voicebox: Text-guided multilingual universal speech generation at scale.
\newblock \emph{Advances in neural information processing systems}, 36:\penalty0 14005--14034, 2023.

\bibitem[Lee et~al.(2025)Lee, Choi, Kim, and Lee]{lee2025hierspeech++}
Sang-Hoon Lee, Ha-Yeong Choi, Seung-Bin Kim, and Seong-Whan Lee.
\newblock Hierspeech++: Bridging the gap between semantic and acoustic representation of speech by hierarchical variational inference for zero-shot speech synthesis.
\newblock \emph{IEEE Transactions on Neural Networks and Learning Systems}, 2025.

\bibitem[Lei et~al.(2025)Lei, Xu, Lin, Zhang, Tan, Chen, Yu, Zhang, Yang, Zhu, et~al.]{lei2025levo}
Shun Lei, Yaoxun Xu, Zhiwei Lin, Huaicheng Zhang, Wei Tan, Hangting Chen, Jianwei Yu, Yixuan Zhang, Chenyu Yang, Haina Zhu, et~al.
\newblock Levo: High-quality song generation with multi-preference alignment.
\newblock \emph{arXiv preprint arXiv:2506.07520}, 2025.

\bibitem[Li et~al.(2019)Li, Liu, Liu, Zhao, and Liu]{li2019neural}
Naihan Li, Shujie Liu, Yanqing Liu, Sheng Zhao, and Ming Liu.
\newblock Neural speech synthesis with transformer network.
\newblock In \emph{Proceedings of the AAAI conference on artificial intelligence}, volume~33, pp.\  6706--6713, 2019.

\bibitem[Li et~al.(2024)Li, Tian, Li, Deng, and He]{li2024autoregressive}
Tianhong Li, Yonglong Tian, He~Li, Mingyang Deng, and Kaiming He.
\newblock Autoregressive image generation without vector quantization.
\newblock \emph{Advances in Neural Information Processing Systems}, 37:\penalty0 56424--56445, 2024.

\bibitem[Liu et~al.(2024)Liu, Liu, Hu, Gao, Zhang, and Ling]{liu2024diffstyletts}
Jiaxuan Liu, Zhaoci Liu, Yajun Hu, Yingying Gao, Shilei Zhang, and Zhenhua Ling.
\newblock Diffstyletts: Diffusion-based hierarchical prosody modeling for text-to-speech with diverse and controllable styles.
\newblock \emph{arXiv preprint arXiv:2412.03388}, 2024.

\bibitem[Meng et~al.(2024)Meng, Zhou, Liu, Chen, Han, Hu, Liu, Li, Zhao, Wu, et~al.]{meng2024autoregressive}
Lingwei Meng, Long Zhou, Shujie Liu, Sanyuan Chen, Bing Han, Shujie Hu, Yanqing Liu, Jinyu Li, Sheng Zhao, Xixin Wu, et~al.
\newblock Autoregressive speech synthesis without vector quantization.
\newblock \emph{arXiv preprint arXiv:2407.08551}, 2024.

\bibitem[Mentzer et~al.()Mentzer, Minnen, Agustsson, and Tschannen]{mentzerfinite}
Fabian Mentzer, David Minnen, Eirikur Agustsson, and Michael Tschannen.
\newblock Finite scalar quantization: Vq-vae made simple.
\newblock In \emph{The Twelfth International Conference on Learning Representations}.

\bibitem[Nishimura et~al.()Nishimura, Hirose, Ohi, Nakayama, and Inoue]{nishimurahall}
Yuto Nishimura, Takumi Hirose, Masanari Ohi, Hideki Nakayama, and Nakamasa Inoue.
\newblock Hall-e: Hierarchical neural codec language model for minute-long zero-shot text-to-speech synthesis.
\newblock In \emph{The Thirteenth International Conference on Learning Representations}.

\bibitem[OpenAudio(2024)]{openaudios1}
OpenAudio.
\newblock Openaudio s1: a cutting-edge text-to-speech model that performs like voice actors.
\newblock \emph{https://openaudio.com/blogs/s1}, 2024.

\bibitem[Pasini et~al.(2024)Pasini, Nistal, Lattner, and Fazekas]{pasini2024continuous}
Marco Pasini, Javier Nistal, Stefan Lattner, and George Fazekas.
\newblock Continuous autoregressive models with noise augmentation avoid error accumulation.
\newblock \emph{arXiv preprint arXiv:2411.18447}, 2024.

\bibitem[Peng et~al.(2024)Peng, Huang, Li, Mohamed, and Harwath]{peng2024voicecraft}
Puyuan Peng, Po-Yao Huang, Shang-Wen Li, Abdelrahman Mohamed, and David Harwath.
\newblock Voicecraft: Zero-shot speech editing and text-to-speech in the wild.
\newblock In \emph{Proceedings of the 62nd Annual Meeting of the Association for Computational Linguistics (Volume 1: Long Papers)}, pp.\  12442--12462, 2024.

\bibitem[Peng et~al.(2025)Peng, Yu, Wang, Chang, Sun, Dong, Zhu, Xu, Bao, Wang, et~al.]{peng2025vibevoice}
Zhiliang Peng, Jianwei Yu, Wenhui Wang, Yaoyao Chang, Yutao Sun, Li~Dong, Yi~Zhu, Weijiang Xu, Hangbo Bao, Zehua Wang, et~al.
\newblock Vibevoice technical report.
\newblock \emph{arXiv preprint arXiv:2508.19205}, 2025.

\bibitem[Ping et~al.(2017)Ping, Peng, Gibiansky, Arik, Kannan, Narang, Raiman, and Miller]{ping2017deep}
Wei Ping, Kainan Peng, Andrew Gibiansky, Sercan~O Arik, Ajay Kannan, Sharan Narang, Jonathan Raiman, and John Miller.
\newblock Deep voice 3: Scaling text-to-speech with convolutional sequence learning.
\newblock \emph{arXiv preprint arXiv:1710.07654}, 2017.

\bibitem[Ren et~al.(2020)Ren, Hu, Tan, Qin, Zhao, Zhao, and Liu]{renfastspeech}
Yi~Ren, Chenxu Hu, Xu~Tan, Tao Qin, Sheng Zhao, Zhou Zhao, and Tie-Yan Liu.
\newblock Fastspeech 2: Fast and high-quality end-to-end text to speech.
\newblock In \emph{International Conference on Learning Representations}, 2020.

\bibitem[Shen et~al.(2018)Shen, Pang, Weiss, Schuster, Jaitly, Yang, Chen, Zhang, Wang, Skerrv-Ryan, et~al.]{shen2018natural}
Jonathan Shen, Ruoming Pang, Ron~J Weiss, Mike Schuster, Navdeep Jaitly, Zongheng Yang, Zhifeng Chen, Yu~Zhang, Yuxuan Wang, Rj~Skerrv-Ryan, et~al.
\newblock Natural tts synthesis by conditioning wavenet on mel spectrogram predictions.
\newblock In \emph{2018 IEEE international conference on acoustics, speech and signal processing (ICASSP)}, pp.\  4779--4783. IEEE, 2018.

\bibitem[Shen et~al.(2023)Shen, Ju, Tan, Liu, Leng, He, Qin, Bian, et~al.]{shen2023naturalspeech}
Kai Shen, Zeqian Ju, Xu~Tan, Eric Liu, Yichong Leng, Lei He, Tao Qin, Jiang Bian, et~al.
\newblock Naturalspeech 2: Latent diffusion models are natural and zero-shot speech and singing synthesizers.
\newblock In \emph{The Twelfth International Conference on Learning Representations}, 2023.

\bibitem[Team et~al.(2025)Team, Xiao, Li, Han, Bai, Cai, Chen, Chen, Cong, Cui, et~al.]{team2025minicpm4}
MiniCPM Team, Chaojun Xiao, Yuxuan Li, Xu~Han, Yuzhuo Bai, Jie Cai, Haotian Chen, Wentong Chen, Xin Cong, Ganqu Cui, et~al.
\newblock Minicpm4: Ultra-efficient llms on end devices.
\newblock \emph{arXiv preprint arXiv:2506.07900}, 2025.

\bibitem[Wang et~al.(2024)Wang, Zeng, Zhang, Ma, Zhu, Cai, Zhao, Jiang, and Chen]{wang2024ham}
Chunhui Wang, Chang Zeng, Bowen Zhang, Ziyang Ma, Yefan Zhu, Zifeng Cai, Jian Zhao, Zhonglin Jiang, and Yong Chen.
\newblock Ham-tts: Hierarchical acoustic modeling for token-based zero-shot text-to-speech with model and data scaling.
\newblock \emph{arXiv preprint arXiv:2403.05989}, 2024.

\bibitem[Wang et~al.(2025{\natexlab{a}})Wang, Liu, Meng, Li, Yang, Zhao, Sun, Liu, Sun, Zhou, et~al.]{wang2025felle}
Hui Wang, Shujie Liu, Lingwei Meng, Jinyu Li, Yifan Yang, Shiwan Zhao, Haiyang Sun, Yanqing Liu, Haoqin Sun, Jiaming Zhou, et~al.
\newblock Felle: Autoregressive speech synthesis with token-wise coarse-to-fine flow matching.
\newblock \emph{arXiv preprint arXiv:2502.11128}, 2025{\natexlab{a}}.

\bibitem[Wang et~al.(2025{\natexlab{b}})Wang, Jiang, Ma, Zhang, Liu, Li, Liang, Zheng, Wang, Feng, et~al.]{wang2025spark}
Xinsheng Wang, Mingqi Jiang, Ziyang Ma, Ziyu Zhang, Songxiang Liu, Linqin Li, Zheng Liang, Qixi Zheng, Rui Wang, Xiaoqin Feng, et~al.
\newblock Spark-tts: An efficient llm-based text-to-speech model with single-stream decoupled speech tokens.
\newblock \emph{arXiv preprint arXiv:2503.01710}, 2025{\natexlab{b}}.

\bibitem[Wang et~al.()Wang, Zhan, Liu, Zeng, Guo, Zheng, Zhang, Zhang, Zhang, and Wu]{wangmaskgct}
Yuancheng Wang, Haoyue Zhan, Liwei Liu, Ruihong Zeng, Haotian Guo, Jiachen Zheng, Qiang Zhang, Xueyao Zhang, Shunsi Zhang, and Zhizheng Wu.
\newblock Maskgct: Zero-shot text-to-speech with masked generative codec transformer.
\newblock In \emph{The Thirteenth International Conference on Learning Representations}.

\bibitem[Wu et~al.(2025)Wu, Deng, Li, Kong, and Lui]{wu2025clear}
Chun~Yat Wu, Jiajun Deng, Guinan Li, Qiuqiang Kong, and Simon Lui.
\newblock Clear: Continuous latent autoregressive modeling for high-quality and low-latency speech synthesis.
\newblock \emph{arXiv preprint arXiv:2508.19098}, 2025.

\bibitem[Xie et~al.(2025)Xie, Shen, Li, Xie, Tang, and Hu]{xie2025fireredtts}
Kun Xie, Feiyu Shen, Junjie Li, Fenglong Xie, Xu~Tang, and Yao Hu.
\newblock Fireredtts-2: Towards long conversational speech generation for podcast and chatbot.
\newblock \emph{arXiv preprint arXiv:2509.02020}, 2025.

\bibitem[Xu et~al.(2025)Xu, Guo, He, Hu, He, Bai, Chen, Wang, Fan, Dang, et~al.]{xu2025qwen2}
Jin Xu, Zhifang Guo, Jinzheng He, Hangrui Hu, Ting He, Shuai Bai, Keqin Chen, Jialin Wang, Yang Fan, Kai Dang, et~al.
\newblock Qwen2. 5-omni technical report.
\newblock \emph{arXiv preprint arXiv:2503.20215}, 2025.

\bibitem[Zhang et~al.(2025)Zhang, Guo, Yang, Yu, Zhang, Lei, Mai, Yan, Yang, Yang, et~al.]{zhang2025minimax}
Bowen Zhang, Congchao Guo, Geng Yang, Hang Yu, Haozhe Zhang, Heidi Lei, Jialong Mai, Junjie Yan, Kaiyue Yang, Mingqi Yang, et~al.
\newblock Minimax-speech: Intrinsic zero-shot text-to-speech with a learnable speaker encoder.
\newblock \emph{arXiv preprint arXiv:2505.07916}, 2025.

\bibitem[Zhou et~al.(2025)Zhou, Zhou, He, Zhou, Wang, Deng, and Shu]{zhou2025indextts2}
Siyi Zhou, Yiquan Zhou, Yi~He, Xun Zhou, Jinchao Wang, Wei Deng, and Jingchen Shu.
\newblock Indextts2: A breakthrough in emotionally expressive and duration-controlled auto-regressive zero-shot text-to-speech.
\newblock \emph{arXiv preprint arXiv:2506.21619}, 2025.

\bibitem[Zhu et~al.(2024)Zhu, Tian, and Xie]{zhu2024autoregressive}
Xinfa Zhu, Wenjie Tian, and Lei Xie.
\newblock Autoregressive speech synthesis with next-distribution prediction.
\newblock \emph{arXiv preprint arXiv:2412.16846}, 2024.

\end{thebibliography}

% \newpage
% \appendix
% \input{appendix}

\end{document}